\documentclass{emulateapj}
\usepackage{multirow}
\usepackage{txfonts}
\usepackage[figuresright]{rotating}
\usepackage{booktabs}

\shorttitle{Star clusters in M31}
\shortauthors{Wang et al.}

\begin{document}

\title{NEW 2MASS NEAR-INFRARED PHOTOMETRY FOR GLOBULAR CLUSTERS IN M31}
\author{Song Wang\altaffilmark{1,2}, Jun Ma\altaffilmark{1}, Zhenyu Wu\altaffilmark{1},
  AND Xu Zhou\altaffilmark{1}}


\altaffiltext{1}{Key Laboratory of Optical Astronomy, National Astronomical Observatories,
Chinese Academy of Sciences, Beijing 100012, China;\\ majun@nao.cas.cn}

\altaffiltext{2}{University of Chinese Academy of Sciences, Beijing 100039, China}

\begin{abstract}
We present 2MASS $JHK_{\rm s}$ photometry for 913 star clusters and candidates in the field of
M31, which are selected from the latest Revised Bologna Catalog of M31 globular clusters (GCs)
and candidates. The photometric measurements in this paper supplement this catalog, and
provide a most comprehensive and homogeneous photometric catalog for M31 GCs in the
$JHK_{\rm s}$ bandpasses. In general, our photometry is consistent with previous measurements.
The globular cluster luminosity function (GCLF) peaks for the confirmed GCs derived by fitting
a $t_5$ distribution using maximum likelihood method are:
$J_0 = 15.348_{-0.208}^{+0.206}$, $H_0 = 14.703_{-0.180}^{+0.176}$, and ${K_{\rm s}}_0 = 14.534_{-0.146}^{+0.142}$,
all of which agree well with previous studies.
The GCLFs are different between metal-rich (MR) and metal-poor
(MP), inner and outer subpopulations, as that MP clusters are fainter than their MR counterparts,
and the inner clusters are brighter than the outer ones, which confirm previous results.
The NIR colors of the GC candidates are on average redder than those of the confirmed GCs,
which lead to an obscure bimodal distribution of the color indices.
The relation of $(V-K_{\rm s})_0$ and metallicity shows a notable
departure from linearity, with a shallower slope towards the redder end.
The color-magnitude diagram (CMD) and color-color
diagram show that many GC candidates are located out of the evolutionary tracks, suggesting
that some of them may be false M31 GC candidates.
The CMD also shows that the initial mass function of M31 GCs covers a large range,
and the majority of the clusters have initial masses between $10^3$ and $10^6$ $M_{\odot}$.

\end{abstract}


\section{INTRODUCTION}
\label{intro.sec}

Globular clusters (GCs) provide a unique laboratory for investigating the formation and
evolution of their host galaxies. The form of the mass spectrum for GCs is nearly identical to
the mass function of the parent molecular cloud cores \citep{mp96}. The brightness
distribution of GCs, known as globular cluster luminosity function (GCLF),  can be used to
constrain possibilities for GC formation and destruction \citep[see][and references
therein]{nantais06}.

By virtue of the natural advantage of being located at a reasonable distance, nearby galaxies
(especially M31) offer us an ideal environment for detailed studies of cluster systems. A
large number of GCs have been identified in M31 since \citet{hubble32}, and the latest Revised
Bologna Catalog of M31 GCs and candidates \citep[hereafter RBC V.5,][]{gall04,gall06,gall09},
which is a compilation of photometry and identifications from many previous catalogs,
published 625 confirmed GCs and 331 GC candidates. Based on the growing number and updated sample
of M31 GCs and candidates, many works have probe the GCLF in M31 in detail.
\citet{crampton85} found that the mean luminosity of GCs in M31 decreases markedly with
increasing galactocentric distance, indicating that clusters with larger projected distances
are fainter than those closer to the nucleus \citep{gnedin97}.
\citet{og97} obtained distance moduli to M31 and M87 from GCLF by applying the corrections for
dynamical evolution, and found surprising consistency
of the predicted and observed distances, which confirmed the GCLF as a distance indicator.
These authors also concluded that the mass functions of GCs in the Milky Way (MW), M31, and M87
were universal at the birth of these systems, although spanning a wide range of masses.
\citet{bhb01} measured the LF for M31 GCs, and found that inner clusters have a GCLF peak brighter
than the outer ones, while the metal-rich (MR) clusters are brighter than their metal-poor (MP)
counterparts. The variation in the M31 GCLF seems to be due to various factors: metallicity, age,
cluster initial mass function (IMF), and dynamical destruction \citep{bhb01}.
Recently, \citet{nantais06} compared the GCLFs of the MW, M31, and the Sculptor Group spiral
galaxies, and found that the GCLF of the MW is consistent with that of M31.

Near-infrared (NIR) colors can help to distinguish among star formation histories and IMFs
\citep{Barton03}. The Two Micron
All Sky Survey (2MASS), performed between 1997 June and 2001 February, covers 99.998\% of the
celestial sphere in the $J$ (1.25 $\mu$m), $H$ (1.65 $\mu$m), and $K_{\rm s}$
(2.16 $\mu$m) bandpasses \citep{Skrutskie06}. These observations were conducted with two 1.3 m
diameter telescopes located at Arizona and Chile. The 2MASS All-Sky Data Release contains the
observation of M31 with an integration time of 7.8 s for each exposure, while a new 2.8 deg$^2$
NIR survey from the 2MASS 6X program across the extent of the optical disk of M31, with an
exposure time of 6 times the normal exposure, provides a clearer view of the galaxy center.
\citet{gall04} identified 693 known and candidate GCs in M31 using the 2MASS database, and
derived their 2MASS $JHK_{\rm s}$ magnitudes. After adding the mean
difference between the 2MASS photometry and previous NIR photometry, the newly assembled NIR
dataset were implemented into a revised version of Bologna Catalog.
These authors also showed that the $V-K_{\rm s}$ color provides a powerful tool to discriminate
between M31 clusters and background galaxies.
\citet{santos13} presented 2MASS photometry and color for a sample of Local Group clusters
younger than $\sim$100 Myr, and found that the embedded clusters, which are heavily obscured
by dust, generally have a redder $H-K_{\rm s}$ color than older ones, from which gas and
dust have already been ejected. These authors also concluded that the brightest clusters can
be split into young and old subsamples from $H-K_{\rm s}$ color. Considering that a more
extended and homogenous photometry in the NIR is important for the studies on M31 star
clusters, we would carry out new photometry for them using 2MASS images.

In this paper, we provide NIR photometry for a set of star clusters in M31 using images from
2MASS. This paper is organized as follows. In Section 2 we present the sample selection,
$JHK_{\rm s}$ photometry, and comparisons with previous measurements.
A discussion on the properties of the sample clusters will be given
in Section 3, including the GCLF and color distributions. Finally, we will summarize our
results in Section 4.

\section{DATA}
\label{data.sec}

\subsection{Sample}
\label{s:samp}
We selected our sample GCs and candidates from RBC V.5,
which contains precise cluster coordinates, classifications,
metallicities, reddening values, radial velocities, galactocentric distances, structure
parameters, and photometry including {\sl GALEX} (FUV and NUV), optical broad-band, and
2MASS NIR magnitudes for 2060 objects. However, RBC V.5 provided $JHK_{\rm s}$ magnitudes in
three bands only for $\sim$1100 objects. In addition, there were no magnitude errors for
$JHK_{\rm s}$ bands in RBC V.5. Therefore, homogeneous photometric data and precise magnitude
errors for $JHK_{\rm s}$ bands are urgently needed. We selected classes 1, 2, 3, and 8 from
column `f' in RBC V.5, which include GCs, candidate GCs, controversial objects, and
extended clusters. This resulted in an initial selection of 991 objects.
We first employed {\sc iraf/daofind} to find the sources in the images and matched them to
the coordinates of these objects given by RBC V.5, and then checked each object visually in the
images to confirm the positions for the photometry.
During this process, we found that there are 68 clusters
for which accurate photometric measurements cannot be obtained with different reasons.
Some clusters (B024D, B053D, B061D, B523, B537, BH16, M023, SK068B, SK110A) are superimposed onto
a bright or strongly variable background. Some clusters (B102D, HEC2, MCEC4) are very faint and
the signal-to-noise ratio (S/N) is low. Some clusters (B189D, B246, B274, B538, BH09, BH25, C011,
C041, KHM31-77, M050, M065, M075, M078, NB89, SK059B, SK061A, SK063C, SK097B, SK104B, SK153C) are
very close to other objects. Some clusters (B016D, B088D, B530, BH01, BH03, BH04, BH28, DAO93, H20,
HEC6, HEC9, HEC13, KHM31-345, M055, M079, NB17, SH06, SK049A, SK222C, SK216B,
SK223B, SK242B) are indiscernible with the coordinates presented by RBC V.5 in the 2MASS images.
There is one or two nearby objects within 2$\sim$4 pixels for five clusters (B016D, B530,
M055, SH06, SK223B) of them, however, after careful comparison
with the images from Local Group Survey \citep[LGS;][]{massey06} and from Digitized Sky Survey (DSS),
these objects cannot be confirmed due to their low S/N values.
In addition, we found that some clusters (B196D, B257D, B318, B368, B453, B475, G099, KHM31-85, M056,
M068, M088, M089, SK166B, V031) have one or more
nearby bright objects in the images from LGS or from Wide Field Planetary
Camera on {\it Hubble Space Telescope}, indicating that they are ``adhered'' in 2MASS images
because of the poor resolution.
\citet{ma12b} presented $JHK_{\rm s}$ photometry, ages, and masses for 10 newly-discovered
halo GCs in M31. So, we will not re-estimated $JHK_{\rm s}$ magnitudes for these GCs.
Finally, here we will derive NIR photometry for the remaining 913 star clusters and
cluster candidates.
The position of cluster H26 presented by RBC V.5 may be of insufficient accuracy,
with R.A. and decl. being 00:59:27.37 and +37:41:34.1 \citep{huxor08}.
We checked the images from 2MASS and from DSS, and corrected the coordinate to
00:59:27.48 and +37:41:31.37. The updated position was then used for the following photometry.
Figure 1 shows the spatial distribution of the sample clusters in M31.

\begin{figure}[!htb]
\figurenum{1}
\center
\resizebox{\hsize}{!}{\rotatebox{0}
{\includegraphics{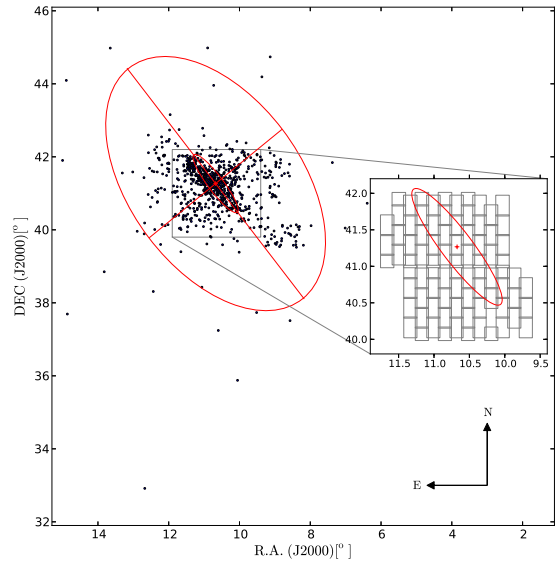}}}
\caption{Location of our sample star clusters in relation to M31. The inner red ellipse
has a radius of $R =1^{\circ}$ and an inclination angle of $i = 77.5^{\circ}$,
while the outer red ellipse has a radius
of 55 kpc and is flattened to $b/a = 0.6$, as given in \citet{rich09}. The filled circles
represent the sample clusters in this paper. The subplot presents the used images from
2MASS-6X survey, and the red ellipse has the same dimension of $R =1^{\circ}$.}
\label{fig1}
\end{figure}

\subsection{Integrated Photometry}
\label{phot.sec}

We used the 2MASS archival images of M31 in the $JHK_{\rm s}$ bands to do photometry. The
images were retrieved using the 2MASS Batch Image
Service\footnote{http://irsa.ipac.caltech.edu/applications/2MASS/IM/batch.html.}. Considering
that the 2MASS-6X images, which were observed with 6 times the normal exposure of 7.8 s, have a
deeper magnitude limit and a higher S/N than the All-Sky Release
Survey images, the 2MASS-6X data were preferred. In this paper, photometry for 627 objects
were performed using the 2MASS-6X data, while photometry for the rest 286 ones were done using
the All-Sky Release Survey data. The uncompressed atlas images provided were used, with a
resampled spatial resolution of $\sim$$1''$ pixel$^{-1}$. We performed aperture photometry of
the 913 M31 star clusters in all of the $JHK_{\rm s}$ bands to provide a comprehensive and
homogeneous photometric catalog for them. The photometry routine we used is {\sc iraf/daophot}
\citep{stet87}.

To determine the total luminosity of each object, we produced a curve of growth from $J$-band
photometry obtained through apertures with radii in the range of 4-16 pixel with 1 pixel
increments. The most appropriate photometric radius from the curve that encloses the total
cluster light but excludes lights from extraneous field stars were adopted. In addition, we
checked the images of all the three bands for each cluster by visual examination to make sure
that the photometric aperture was not so large as to be contaminated by other sources.
The local sky background was
measured in an annulus with an inner radius 1 pixel larger than the photometric radius and 5
pixels wide. The sky fitting algorithm was set to ``mode''.
The derived magnitudes with large uncertainties ($> 1$ mag) were abandoned here to keep high
photometry precision.
The instrumental magnitudes were then calibrated using the relevant zero points
obtained from the photometric header keywords of each image. Figure 2 shows the aperture
distribution as a function of $J$-band magnitude, with a highest peak at $4''$. In general,
magnitudes of brighter objects are measured with larger apertures.

\begin{figure}[!htb]
\figurenum{2}
\center
\resizebox{\hsize}{!}{\rotatebox{0}
{\includegraphics{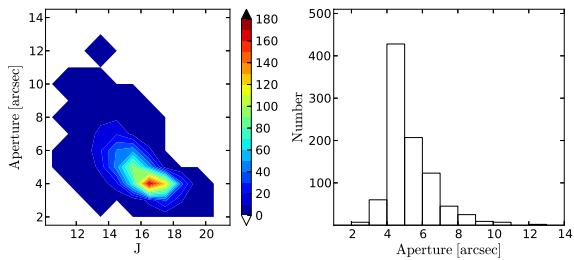}}}
\vspace{0.1cm}
\caption{($Left~panel$) Contour plot of the aperture as a function of $J$-band magnitude.
($Right~panel$) Aperture distribution for M31 star clusters.}
\label{fig:fig2}
\end{figure}

Finally, except for 9 objects in the $J$ band, 31 objects in the $H$ band, and 51 objects in
the $K_{\rm s}$ band, we obtained photometry for 913 star clusters in the individual
$JHK_{\rm s}$ bands.
Table 1 lists our new $JHK_{\rm s}$ magnitudes and the aperture radii used,
in conjunction with the magnitude uncertainties given by
{\sc daophot}. Figure 3 shows the uncertainty distribution as a function of magnitude for the
$JHK_{\rm s}$ bands. Filled circles represent photometry performed on the 2MASS-6X images,
while red pluses represent photometry performed on the 2MASS All-Sky images. It is clear that
the photometry from 2MASS-6X data show smaller magnitude uncertainties, which are due to high
S/N with longer exposures.
Objects in these dashed boxes are bright sources with slightly large magnitude uncertainties
($> 0.2$ mag). Six clusters (AU008, AU010, B132, NB21, NB39, NB41) have
large magnitude uncertainties in all three bands, two clusters (B070, B119) have large magnitude
uncertainties in $K_{\rm s}$ band, two clusters (B104, NB108) have large magnitude
uncertainties in $JH$ bands,  and three clusters (B264, B324, SK050A) have large magnitude
uncertainties in $HK_{\rm s}$ bands. All these clusters are close to the galaxy center and suffer
from a bright background contamination.

\begin{figure}[!htb]
\figurenum{3}
\center
\resizebox{\hsize}{!}{\rotatebox{0}
{\includegraphics{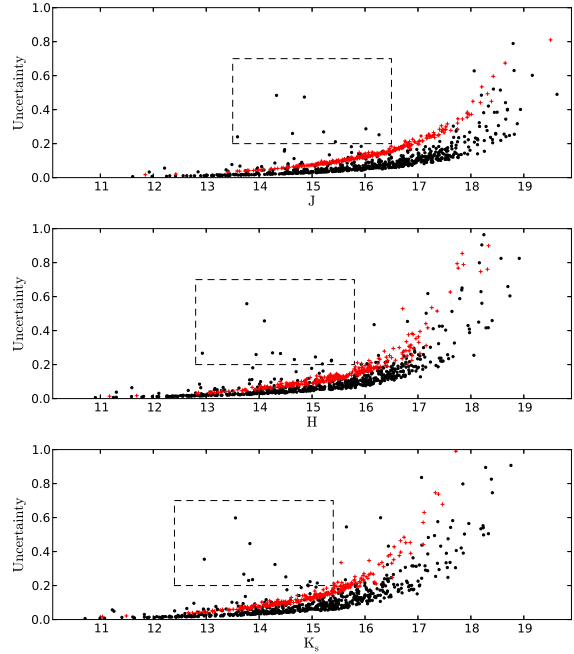}}}
\caption{Distribution of uncertainty as a function of magnitude for the $JHK_{\rm s}$ bands.
The filled circles represent photometry performed on the 2MASS-6X images,
while the red pluses represent photometry measured on the 2MASS All-Sky images.
Objects in these dashed boxes are bright sources with slightly large magnitude uncertainties.}
\label{fig:fig3}
\end{figure}

\subsection{Comparison with Previously Published Photometry}
\label{compare.sec}

To examine the quality and reliability of our photometry, we compared the aperture magnitudes
of the 913 objects obtained here with previous results, in the sense of values from
this paper minus values from the others.
Figure 4 and Figure 5 plot the comparison of our obtained 2MASS photometry with the aperture
photometry from 2MASS catalog, with  $r=4''$ for the Point Source Catalog (PSC) and $r=5''$
for the Extended Source Catalog (XSC) following \citet{gall04}.
The 2MASS PSC was derived by combining the 2MASS All-Sky PSC
and the 2MASS 6X Point Source Working Database (6X-PSWDB), with the 6X-PSWDB being preferred.
The 2MASS XSC was derived by combining the 2MASS All-Sky XSC
and the 2MASS 6X Extended Source Working Database (6X-XSWDB), with the 6X-XSWDB being preferred.
Each panel displays the mean difference ($\Delta$), rms scatter ($\sigma$),
and the number points ($N$) for the comparison.
There were nine objects with larger offset ($| \Delta m |> 1$ mag) for the comparison with PSC,
as listed in Table 2. It was surprised that the magnitudes of eight of these clusters were
measured with an aperture of $r=4''$ here, which is the same as adopted in the PSC. We checked the images
for these clusters and found that some clusters (SK049C, SK054C, SK086C) are
located in a strongly variable background, while some clusters (B038D, B104, SK047B) lie
close to the M31 center. In addition, there is one very bright source near SK115B.
For objects in the PSC, the sky background is measured in an annulus with an
inner radius of $14''$ and an outer radius of $20''$. To check whether the discrepancy is caused
by the different choice of the annulus for background, we used the same annulus with PSC and
re-estimated magnitudes for the eight clusters. We found that most of the newly derived magnitudes
agree well with those from PSC, except SK115B, and SK136C.
For SK115B, about a quarter of the area of the annulus ($14''$-$20''$) is affected by the nearby
bright source, which would lead to a high estimate of the sky background.
For SK136C, the $J$-band magnitude given by PSC is 13.419, much brighter than the $H$-band
magnitude 16.700, which is suspicious.
We think that the annulus set in this paper is more reasonable for these clusters.
In addition, for cluster B072D, a larger radius would include lights from outer sources.
In this paper, photometry was performed with an aperture radius larger than $4''$ for more than
400 clusters, which may lead to brighter magnitudes than those in 2MASS PSC, on the average.
We concluded that this is the main reason for the offset of the $JHK_{\rm s}$ magnitudes
($-0.029$, $-0.051$, and $-0.060$).
There is no big difference between the magnitudes measured here and those from the XSC.

\begin{figure}[!htb]
\figurenum{4}
\center
\resizebox{\hsize}{!}{\rotatebox{0}
{\includegraphics{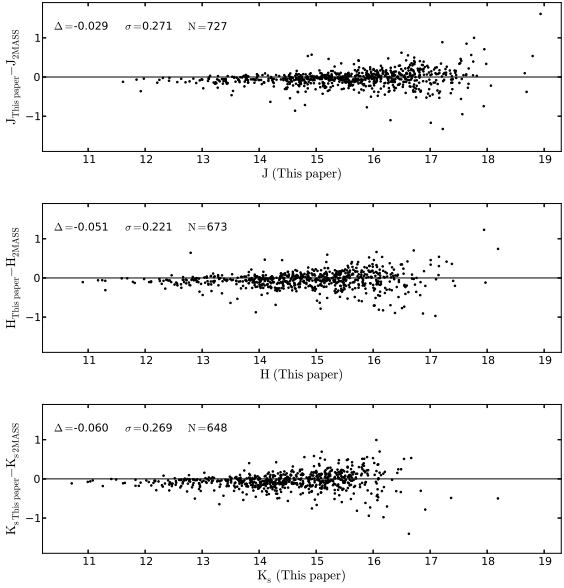}}}
\caption{Comparison of our obtained 2MASS photometry with previous measurements from 2MASS PSC.
The $\Delta$, $\sigma$, and $N$ represent the mean difference, rms scatter, and
the number points used for the comparison, respectively.}
\label{fig:fig4}
\end{figure}

\begin{figure}[!htb]
\figurenum{5}
\center
\resizebox{\hsize}{!}{\rotatebox{0}
{\includegraphics{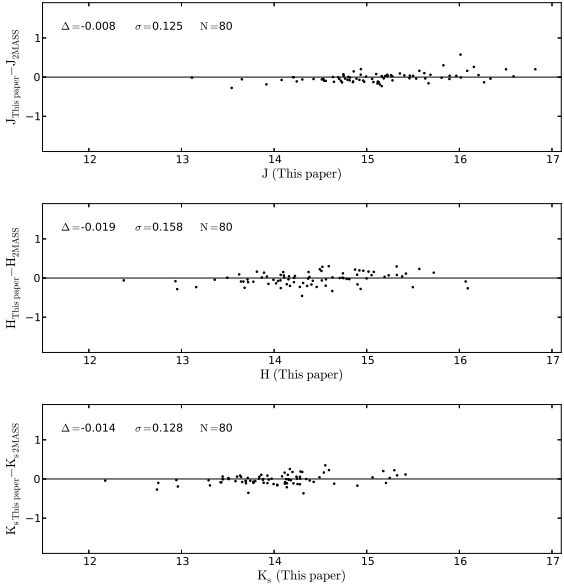}}}
\caption{Comparison of our obtained 2MASS photometry with previous measurements from 2MASS XSC.
The numbers are as in Fig. 4.}
\label{fig:fig5}
\end{figure}

\citet{bh00} measured $JK$ magnitudes for a sample of M31 GCs,
with most of the new NIR data taken with the STELIRCam on the 1.2 m telescope at
Fred Lawrence Whipple Observatory (FLWO), and the observation data for four objects from the 2MASS
\citep{skrutskie97} scans of M31.
\citet{bhb01} obtained NIR photometry for 38 GCs and candidates from
$JHK'$ observation using the Gemini on the Lick Observatory 3 m telescope
and $JK$ observation using the STELIRCam on the FLWO 1.2 m, and updated magnitudes for some
clusters studied in \citet{bh00}. Figure 6 plots the comparison of our obtained 2MASS
photometry with the magnitudes from \citet{bh00, bhb01}.
The average differences for the $J$ and $K_{\rm s}$ bands are $0.024$ and $-0.032$,
which may be caused by different photometry methods, with
the rms scatters around the mean being 0.193 and 0.231.
There is only one cluster (DAO69) with larger magnitude offsets, as listed in Table 2.
A larger radius would include lights from several outer sources,
so we think our photometry result is more reasonable for DAO69.

\begin{figure}[!htb]
\figurenum{6}
\center
\resizebox{\hsize}{!}{\rotatebox{0}
{\includegraphics{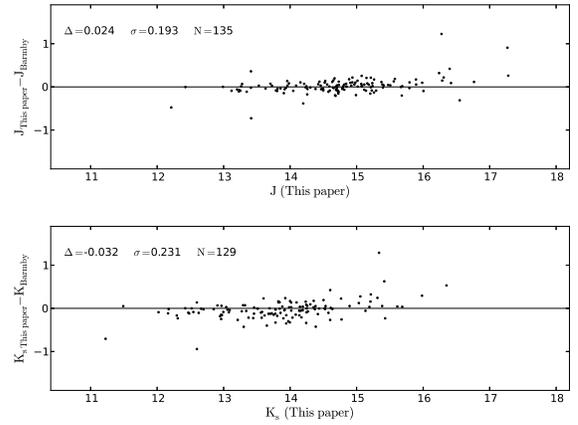}}}
\vspace{0.1cm}
\caption{Comparison of our obtained 2MASS photometry with previous NIR measurements from
\citet{bh00, bhb01}. The numbers are as in Fig. 4.}
\label{fig:fig6}
\end{figure}

Figure 7 displays the comparison of our obtained 2MASS photometry with the magnitudes from RBC
V.5. The 2MASS photometry in RBC V.5 \citep{gall04} was derived from the
2MASS All-Sky PSC and XSC, by correcting the systematic shifts between the magnitudes from the
2MASS catalogs and previous NIR photometries.
In RBC V.5, the 2MASS $JHK_{\rm s}$ magnitudes were transformed to CIT photometric system
\citep{gall04}. However, we needed the original 2MASS $JHK_{\rm s}$ data to compare with our
results, so we reversed the transformation using the equations given by \citet{Carpenter01}.
The average differences for the three bands are $0.059, 0.050$ and $0.032$, with
the rms scatters around the mean being 0.309, 0.264 and 0.327.
There is an obvious offset between their estimated $J$ magnitudes and ours,
which is mainly caused by some large scatters listed in Table 2.
If these clusters with larger offset ($| \Delta m |> 1$ mag) are not included,
the systematic offset in $J$ band can be reduced to be 0.035 mag.
The clusters with the
largest discrepancy are B041, B090, and SK054C.
There are several bright sources beyond
B041, and a larger radius would include lights from these sources.
The $K_{\rm s}$ magnitude of B090 in RBC V.5 is 17.881,
which is from the 2MASS All-Sky PSC.
However, in the 2MASS 6X-PSWDB, the magnitude is 15.308.
We checked the images of the 2MASS All-Sky survey, and found that
the low S/N is the main reason for the $K_{\rm s}$ magnitude inaccuracy.

\begin{figure}[!htb]
\figurenum{7}
\center
\resizebox{\hsize}{!}{\rotatebox{0}
{\includegraphics{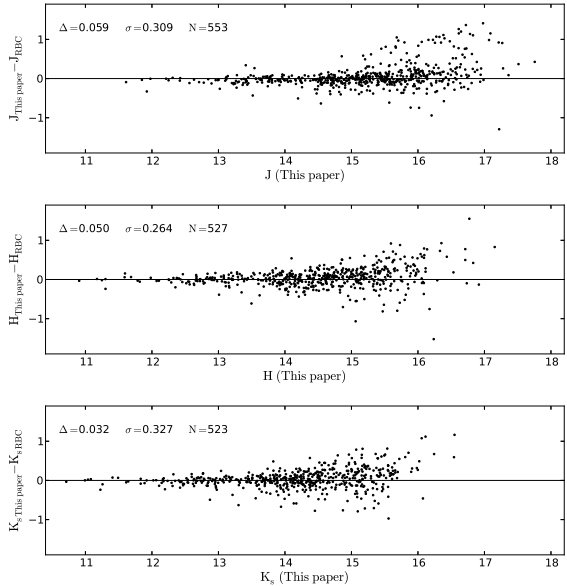}}}
\caption{Comparison of our obtained 2MASS photometry with previous NIR measurements from RBC
V.5. The numbers are as in Fig. 4.}
\label{fig:fig7}
\end{figure}

The four objects with magnitude scatters larger than 2.0 mag are plotted in Figure 8,
in which the circles are photometric apertures adopted here.
As discussed above, the inaccuracy of the $K_{\rm s}$ photometry for B090 is mainly caused by
the low S/N.
B104 lies close to the M31 center, while SK054C is located in a variable background.
The different annulus set for the sky estimation results in the discrepancy of the magnitudes,
and we think that the photometry method in this paper is more reasonable for these clusters.
For SK136C, the $J$ magnitude in 2MASS PSC (13.419), which is much brighter than $H$
magnitude (16.700), may be a typing error.
In all comparisons the average absolute difference is $\leq 0.06$ mag, and the rms scatter
around the mean varies between 0.1 and 0.4.
No system offset between our magnitudes and previous determinations can be seen.

\begin{figure}[!htb]
\figurenum{8}
\center
\resizebox{\hsize}{!}{\rotatebox{0}
{\includegraphics{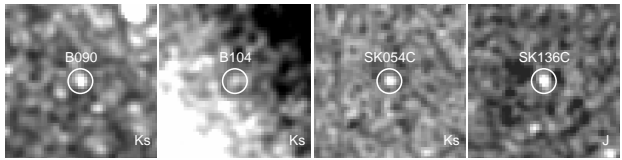}}}
\vspace{0.1cm}
\caption{Images of four star clusters, of which the magnitude
scatters between this paper and other previous studies are larger than 2.0 mag.
The circles are photometric apertures adopted in this paper.}
\label{fig:fig8}
\end{figure}

\subsection{Extinction Correction and Metallicity}
\label{extinction.sec}

\citet{cald09, cald11} published reddening values for a large sample of clusters in M31 by
comparing the observed spectra with model ones. Considering the high accuracy of their
spectral results, we preferentially adopted \citet{cald09, cald11} reddening values.
However, the mode value of 0.13 and values from \citet{bh00} in \citet{cald09, cald11} were
not adopted.
For star clusters with no reddening value given by \citet{cald09, cald11}, we adopted the
results from \citet{bh00}, \citet{fan08}, and \citet{fan10}. \citet{bh00} and \citet{fan08}
determined the reddening for M31 clusters using correlations between optical and infrared
colors and metallicity by defining various ``reddening-free'' parameters. Because the
reddening values from \citet{fan08} comprise a homogeneous data set and the number of GCs
included is greater than that of \citet{bh00}, we preferentially adopted \citet{fan08} reddening
values, followed by those of \citet{bh00}. Finally, the results from \citet{fan10}, which
derived reddening values from spectral-energy distribution fitting, were adopted.

There are 378 star clusters with no available reddening values in the literatures, and we would
calculate reddening values for them in a similar way as described in \citet{kang12}. For clusters
with a de-projected galactocentric distance $<$ 22 kpc, the reddening values were measured as the
mean reddening values of star clusters located within an annulus at every 2 kpc radius from
the center of M31. While for clusters
beyond a de-projected distance of 22 kpc, most of which are located in halo regions,
the reddening value of $E(B-V) = 0.13$ mag was adopted as \citet{kang12} suggested.

\citet{cald11} provided new homogeneous estimates of metallicity for more than 300 GCs in
M31, using high-quality spectra obtained with the Hectospec multifiber spectrograph on the 6.5
m MMT. \citet{kang12} derived mean value of metallicities for a catalog of clusters from previous
literatures \citep{bh00,per02,gall09,cald11}. Metallicities for 312 and 289 clusters in our
sample were given by \citet{cald11} and \citet{kang12}, respectively. Table 1 lists the
metallicities from \citet{cald11} and \citet{kang12} for the sample clusters.

\section{DISCUSSION}
\label{discussion.sec}

We combined the photometry results of $JHK_{\rm s}$ bands newly derived here with those
derived by \citet{ma12b} for 10 GCs in the M31 halo to construct a more comprehensive sample
of 923 clusters in M31 to discuss the properties for them.
The reddening values for the 10 GCs were derived from \citet{ma12b}, while the metallicities
from \citet{cald11} and \citet{kang12} were used for them, respectively.
The $V$-band magnitudes of the
sample clusters were derived from RBC V.5 for the following analysis about the color
distribution. All these magnitudes have been extinction corrected.

\subsection{Luminosity Function}
\label{LF.sec}

To obtain the M31 GCLF peaks, we used least square and maximum likelihood methods to fit a
Gaussian and Student's t distribution on the dereddened $JHK_{\rm s}$ data.
The Gaussian distribution is given as
\begin{equation}
f(x)={\frac{1}{\sqrt{2\pi}\sigma_{G}}}
{\rm exp}(-{\frac{(x-\mu_{G})^2}{2\sigma_{G}^2}})
\end{equation}
The Student's t distribution is defined by
\begin{equation}
f(x)={\frac{\Gamma((n+1)/2)}{\sqrt{\pi~n}\Gamma(n/2)\sigma_{t}}}
(1+{\frac{(x-\mu_{t})^2}{n\sigma_{t}^2}})^{-(n+1)/2}
\end{equation}
where n is the degree of freedom (DOF).
\citet{secker92} and \citet{secker93} reported that the Student's t distribution with DOF of 5
(hereafter ``$t_5$''), which presents a power-law fall-off in its wings, is more robust than the
Gaussian in describing the outlying data points and estimating the GCLF peak. The ``$t_5$''
function is evaluated as
\begin{equation}
f(x)={\frac{8}{3\sqrt{5}\pi\sigma_{t}}}
(1+{\frac{(x-\mu_{t})^2}{5\sigma_{t}^2}})^{-3}
\end{equation}
Some other distributions \citep{baum95, Larsen01} were also investigated for the GCLF.
\citet{baum95} found that a composite of two exponentials is a better fit over the Gaussian
or Student's t distribution to the combined LF of the Galactic and M31 GCs, since those two
distributions fail to deal with the asymmetry and with the sharpness of the peak of the histogram.

Figure 9 displays the luminosity histograms and the best fitting lines for the sample
clusters. The top panels show the fitting with least square technique, while the bottom panels
show the fitting with maximum likelihood method.
The black lines show the fitting with a $t_5$ distribution, while the red lines show the
fitting with a Gaussian distribution.
The maximum likelihood method is used to estimate the most probable parameter values from the
sample data \citep{secker92}, which would not suffer from the effect of binning.
As shown in Figure 9, the fitting with a $t_5$ distribution shows a more extended wing than
that with a Gaussian distribution \citep{secker92}.
Table 3 lists the LF parameters for the sample clusters, the confirmed GCs, and those confirmed GCs
with metallicities available from \citet{kang12}.
The LF peaks for confirmed GCs are much brighter than those for all clusters,
indicating that a clean sample of GCs is critical to obtain accurate LF parameters.

The GCLF peaks derived by \citet{bhb01} are $J_0 = 15.26$ and $K_0 = 14.45$,
by fitting a $t_5$ distribution using the MAXIMUM program written by J. Secker
\citep[see][for details]{secker92}. \citet{nantais06} derived similar results for the
$JHK_{\rm s}$ bands using same methods: $J_0 = 15.31$, $H_0 = 14.76$, and ${K_{\rm s}}_0 = 14.51$,
and these authors concluded that the little fainter peaks than those from \citet{bhb01} could be
real, due to a larger and possibly deeper sample. In this paper, the LF peaks for the confirmed GCs,
derived by fitting a $t_5$ distribution using maximum likelihood method, are:
$J_0 = 15.348_{-0.208}^{+0.206}$, $H_0 = 14.703_{-0.180}^{+0.176}$, and ${K_{\rm s}}_0 = 14.534_{-0.146}^{+0.142}$,
all of which agree well with previous results.
The $\sigma$ parameters derived here are slightly larger than previous studies.

\begin{figure}[!htb]
\figurenum{9}
\center
\resizebox{\hsize}{!}{\rotatebox{0}
{\includegraphics{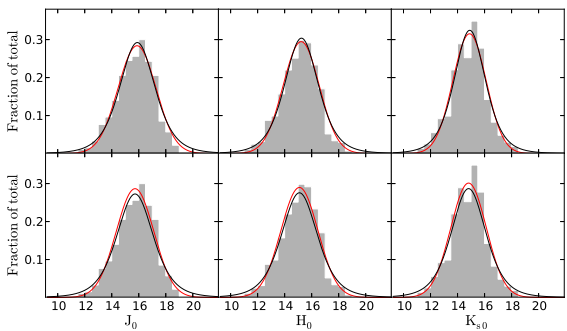}}}
\vspace{0.1cm}
\caption{The luminosity histograms and the best fitting models for the sample clusters.
The top panels show the fitting with least square technique, while the bottom panels
show the fitting with maximum likelihood method. The black and red lines show the
fitting with a $t_5$ and Gaussian distribution, respectively.}
\label{fig:fig9}
\end{figure}

It has long been known that the peak of the GCLF is nearly constant in different galaxies,
providing a standard indicator for the cosmological distance measurement \citep{racine68,
hanes77, ferrarese00}. However, several studies \citep{crampton85, gnedin97, bhb01} presented
that GCLF varies between MR and MP, inner and outer subsamples within
a galaxy, and GCLF peak becomes fainter as the local density of galaxies increases \citep{bt96}.
\citet{Larsen01} found that the $V$-band turnover of the blue GCs is brighter than that of the
red ones by about 0.3 mag on the average, with a study of GCs in 17 nearby early-type galaxies.
\citet{bhb01} found that MR clusters are brighter than MP ones in M31, and inner clusters are
brighter on average than outer clusters, indicating that the luminosity function is different
among these subpopulations.
\citet{Goudfrooij04} confirmed that the GC system in the early-type galaxy NGC 1316, which is
an intermediate-age merger remnant, can be divided into a blue GC subpopulation, consistent
with a Gaussian LF, and a red GC component with a power law LF. These authors also found that
the LF of the inner half of the MR population differs significantly from that of the
outer half.
The difference in GCLF between dwarf and giant ellipticals has been studied by many
authors \citep{harris91, durrell96, strader06, jordan06, ml07}, however, whether there
is a trend of the GCLF peak with the galaxy luminosity is still under debate.

To investigate the effects of metallicity and galactocentric distance on the LF, we divided
the confirmed GCs into four subsamples as \citet{bhb01} did.
The MR and MP subpopulations are divided at [Fe/H] $=-1.1$, using metallicities from \citet{kang12},
while the inner and outer subsamples are divided at $R_{\rm gc}=10$ kpc.
Figure 10 displays the $J_0$ luminosity histograms for the four subsamples,
together with the best-fit $t_5$ distribution using the maximum likelihood method.
Table 4 lists the GCLF fit results in the $JHK_{\rm s}$ bands for the four subsamples.
It is evident that the MP clusters are fainter than their MR counterparts, and the inner clusters
are brighter than the outer ones.
\citet{gnedin97} reported that the destruction of the inner faint, low-density clusters due to
strong tidal shocks may lead to the difference between the LF of the inner and outer
populations.
\citet{kh97} asserted that the difference in the GCLF of the inner and outer halo populations
is due to dynamical evolution and/or a dependence of GCLF shape on the environment.
\citet{og97} found that the predicted differences between the peaks of the inner and
outer cluster populations, due to the tidal shocks and dynamical friction,
agree quantitatively with the observed differences within the errors in the MW, M31, and M87.
However, \citet{bhb01} concluded that metallicity, age, and cluster IMF may also be important
for the variation in the M31 GCLF except dynamical destruction.

\begin{figure}[!htb]
\figurenum{10}
\center
\resizebox{\hsize}{!}{\rotatebox{0}
{\includegraphics{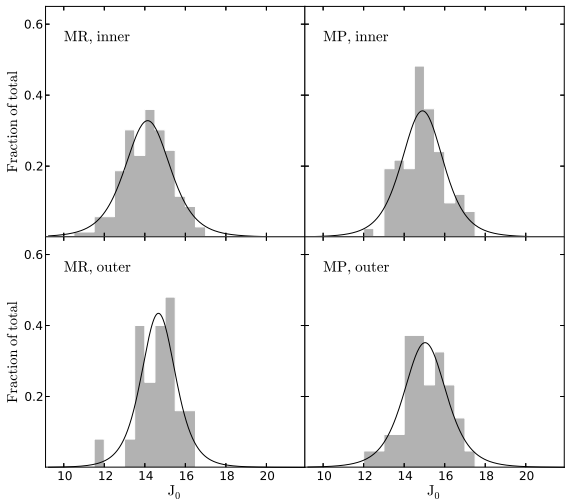}}}
\caption{The $J_0$ luminosity histograms and the best fitting models for four groups of the confirmed
GCs, divided by the galactocentric distances and metallicities from \citet{kang12}.
The black lines show the fitting with a $t_5$ distribution using the maximum likelihood method.}
\label{fig:fig10}
\end{figure}

\subsection{Color Distribution}
\label{color.sec}

Figure 11 displays the NIR color (including $(V-K_{\rm s})_0$ ) distribution of the whole sample
clusters (solid-line histogram) and the confirmed GCs (filled histogram) with magnitude uncertainty
$< 0.5$ mag.
There are some bumps located in the redder wings of the $(J-K_{\rm s})_0$,
$(H-K_{\rm s})_0$, and $(V-K_{\rm s})_0$ colors, which are mainly caused by GC candidates.
The GC systems of many galaxies reveal bimodal optical color distributions
\citep{fbh97, kundu99, gk99, Larsen01, rejkuba01}.
\citet{bh00} have detected the bimodal distribution of the GC colors in M31 using
$(U-V)_0$, $(U-R)_0$, and $(V-K)_0$ with the KMM statistics \citep{mb88, ashman94}.
To check whether the NIR color distributions are bimodal, the KMM algorithm
was also performed here. The homoscedastic fitting, with same variances for
both groups, was assumed. Table 5 lists the parameters returned by the KMM algorithm.
Column (2) and (3) give the estimated mean value and covariance assuming the whole sample as
one group. Column (4) and (5) give the estimated mean value for each group, assuming the whole
sample as two groups, and Column (6) gives the common covariance for them. Column (7) and (8)
give the number points assigned to each group. Column (9) gives the $p$ value, which is an
estimate of the improvement of the two-group fit over a one-group fit, and is interpreted as a
rejection of the single Gaussian model at a confidence level of $1-p$.
The purely NIR color distributions show bimodality at the $\sim$100 per cent confidence level,
however, the size of the redder subsample is very small, and this subsample
is composed mainly of the GC candidates, as shown in Figure 11.
Although Table 5 presents the mean values and number points for the two groups of $(V-K_{\rm s})_0$ color,
these values vary with the initial set,
and the $p$ value is unavailable since the sample is not convergent.
Some previous studies of purely NIR and optical-NIR GC colors in different galaxies have
shown that in some cases the optical color distributions are clearly bimodal,
however, the purely NIR and optical-NIR color distributions are not, or they display
``differing bimodalities'' \citep{blakeslee12}
with those of the optical colors alone \citep[see][and references therein]{cantiello}.

\begin{figure}[!htb]
\figurenum{11}
\center
\resizebox{\hsize}{!}{\rotatebox{0}
{\includegraphics{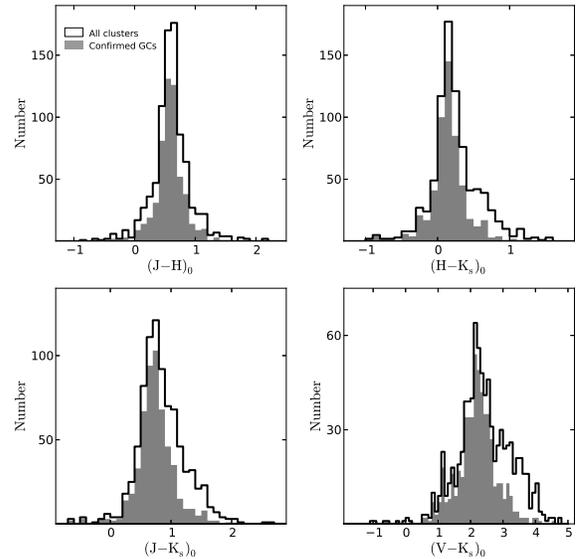}}}
\caption{The color distribution of the sample clusters.}
\label{fig:fig11}
\end{figure}

Figure 12 shows the relation between metallicity and intrinsic colors for the sample clusters
with magnitude uncertainty $< 0.5$ mag. Metallicities in the top panels are from
\citet{kang12}, while metallicities in the bottom panels are from \citet{cald11}.
The open circles represent the confirmed GCs, while the crosses represent the
rest clusters, including candidate GCs, controversial objects, and extended clusters.
The $(V-K_{\rm s})_0$ color index, which is often used as a metallicity indicator,
shows a clearer correlation
with [Fe/H] than other NIR colors, indicating that these NIR intrinsic colors are less sensitive to
metallicity \citep{bh00}. \citet{nantais06} also reported that stellar spectra in the NIR
is much less dependent on metallicity than in the optical.
The relation of $(V-K_{\rm s})_0$ and metallicity shows a notable departure from linearity,
with a shallower slope in the redder part.
The nonlinear correlation of color with metallicity, which are mainly driven by
the horizontal-branch stars, can produce a bimodal color distribution with unimodal metallicity
for a group of old clusters \citep{Yoon06}.

We also use the KMM algorithm to investigate the bimodality of the metallicity distributions for
M31 GCs. Table 6 lists the parameters returned by the KMM algorithm.
The average metallicities for \citet{cald11} and \citet{kang12} are nearly consistent,
while both the two groups
from \citet{kang12} are metal-richer than the two groups from \citet{cald11}. Metallicities
from \citet{kang12} show strong bimodal distribution at the 99.8\% confidence level, while the
hypothesis of a unimodal distribution is rejected only at $\sim$64\% confidence level for
metallicities from \citet{cald11}.
Although many previous studies \citep{ab93,bh00,per02,fan08} have reported that the metallicity
distribution in M31 is bimodal, \citet{cald11} suggested that the metallicity distribution in M31
is not generally bimodal, in strong distinction with the bimodal Galactic globular distribution.

\begin{figure}[!htb]
\figurenum{12}
\center
\resizebox{\hsize}{!}{\rotatebox{0}
{\includegraphics{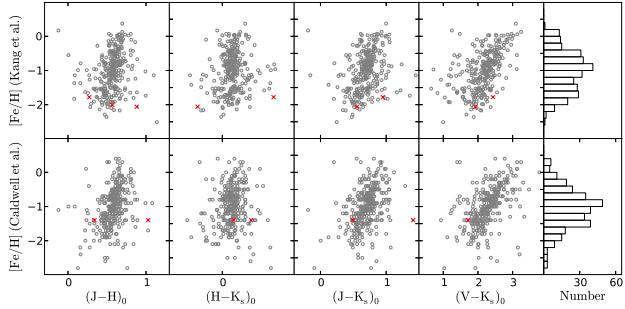}}}
\caption{Metallicity as a function of color. The open circles represent the confirmed GCs,
while the red crosses represent the rest clusters, including candidate GCs,
controversial objects, and extended clusters.}
\label{fig:fig12}
\end{figure}

Figure 13 shows the relation between intrinsic colors and galactocentric distance
$R_{\rm gc}$ for the sample clusters.
Symbols are as in Figure 12.
It can be seen that the colors of candidate GCs are on average redder than those of the confirmed GCs.
We derived the mean color values in different projected galactocentric distance for the
confirmed GCs.
For clusters with galactocentric distance $R_{\rm gc}<$ 30 kpc and $R_{\rm gc}>$ 30 kpc,
the mean values were measured with star clusters located within an annulus at every 3 kpc
and 10 kpc radius from the center of M31, respectively,
all of which were plotted with squares in Figure 13.
\citet{crampton85} found that no radial $(B-V)_0$ color
gradient exists for M31 GCs, however, \citet{sharov88} reported that the $(V-K)_0$ color
shows a weak correlation with the galactocentric distance.
No clear trend is present between the NIR colors and $R_{\rm gc}$ for the confirmed GCs.
It seems that clusters with $R_{\rm gc}$ around 20 kpc
are on average redder than inner clusters in the $(V-K_{\rm s})_0$ color index,
however, this may be caused by the crowding blue clusters around the ``10 kpc ring''
\citep{Gordon06}, which pull the color index down.

\begin{figure}[!htb]
\figurenum{13}
\center
\resizebox{\hsize}{!}{\rotatebox{0}
{\includegraphics{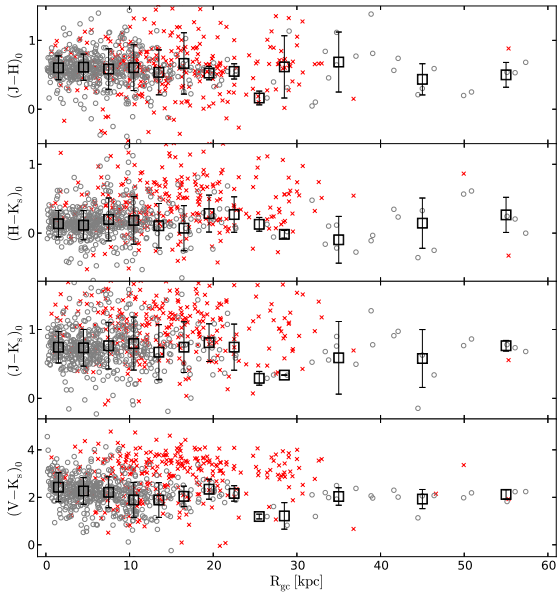}}}
\caption{Intrinsic color as a function of galactocentric distance $R_{\rm gc}$.
Symbols are as in Fig. 12. The squares represent the mean color values in different projected
galactocentric distance for the confirmed GCs.}
\label{fig:fig13}
\end{figure}

The color-magnitude diagram (CMD) provides a qualitative model-independent global indication
of cluster formation history \citep{ma12a, ma13}. Figure 14 displays the CMD of $M_{K_{\rm s}}$ vs.
$(V-K_{\rm s})_0$ for the sample star clusters with magnitude uncertainty $< 0.5$ mag.
Symbols are as in Figure 12.
The absolute magnitudes were derived with the distance modulus of $(m-M)_0=24.47$ with a
distance of $\sim$784 kpc. Several models were added to the CMD to obtain a more detailed
history of cluster formation. Four fading lines
from the simple stellar population (SSP) synthesis model \citet[][hereafter BC03]{bru03}
for a metallicity of $Z=0.004$, $Y=0.24$, assuming a \citet{salp55} stellar IMF, and using
the Padova-1994 isochrones, are plotted on the CMD of M31
star clusters for four different total initial masses: $10^6$, $10^5$, $10^4$, and $10^3$
$M_{\odot}$ (from up to bottom). It seems that many GC candidates in M31 are located out
of the evolutionary tracks.
The majority of M31 GCs fall between these four
fading lines, consistent with previous results \citep{wang10}.

\begin{figure}[!htb]
\figurenum{14}
\center
\resizebox{\hsize}{!}{\rotatebox{0}
{\includegraphics{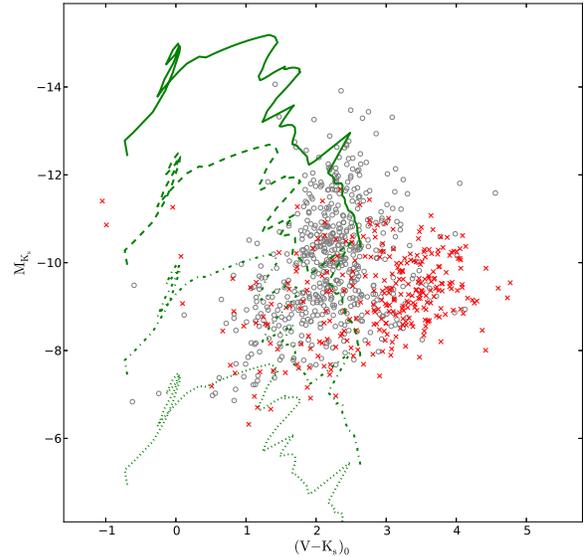}}}
\caption{CMD of the sample clusters. Symbols are as in Fig. 12.
Fading lines are indicated for clusters with total initial masses of $10^6$, $10^5$, $10^4$,
$10^3$ $M_{\odot}$ (from up to bottom), assuming a Salpeter IMF.}
\label{fig:fig14}
\end{figure}

Figure 15 shows the $(V-K_{\rm s})_0$ versus $(J-K_{\rm s})_0$ color-color diagram for M31
star clusters with magnitude uncertainty $< 0.5$ mag. 
Symbols are as in Figure 12.
The theoretical evolutionary paths from
the SSP model BC03 for $Z=0.004$, $Y=0.24$ (black line) and $Z=0.02$, $Y=0.28$ (green line)
are displayed. The horizontal dashed line represents $V-K_{\rm s} = 3$, assuming the reddening value
$E(B-V)=0.13$. As \citet{gall04} reported that, most of the background galaxies have
$V-K_{\rm s} \geq3 $, providing a powerful tool to discriminate between M31 clusters and background
galaxies. It can be seen that a large number of GC candidates are located above the dashed line,
indicating that some background galaxies were mistaken for GC candidates.
However, we should notice that the dashed line is just approximate to the criteria $V-K_{\rm s} = 3$,
since one source with $V-K_{\rm s} > 3$ may be located below the dashed line if the reddening
value is larger than 0.13 mag.
The distribution of the clusters in the color-color diagram are much disperse,
and many GC candidates are located out of the theoretical evolutionary paths,
indicating that many GC candidates may not be true GCs.

\begin{figure}[!htb]
\figurenum{15}
\center
\resizebox{\hsize}{!}{\rotatebox{0}
{\includegraphics{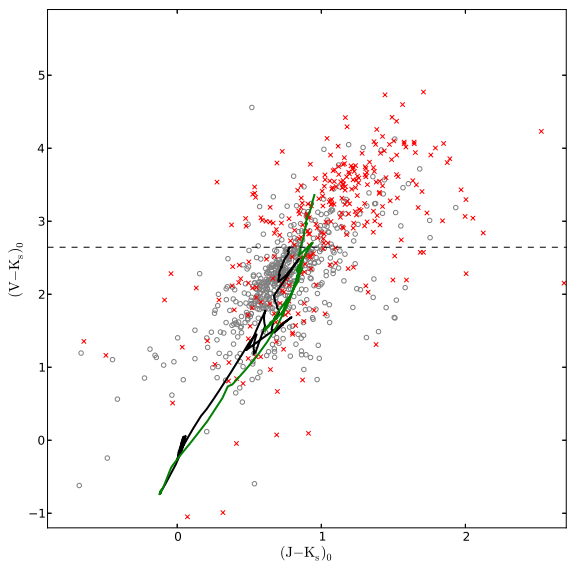}}}
\caption{Color-color diagram of the sample clusters. Symbols are as in Fig. 12.
Theoretical evolutionary paths
for Z = 0.004, Y = 0.24 (black line) and $Z$ = 0.02, $Y$ = 0.28 (green line) are
drawn. The horizontal dashed line represents $V-K_{\rm s} = 3$, assuming the
reddening value $E(B-V)=0.13$.}
\label{fig:fig15}
\end{figure}

\section{SUMMARY}
\label{summary.sec}

In this paper, we performed  $JHK_{\rm s}$ photometric measurements for 913 GCs and
candidates in the field of M31 based on 2MASS images. These sample star clusters
in M31 were selected from RBC V.5, which is the latest Revised
Bologna Catalog of M31 GCs and candidates. The $JHK_{\rm s}$ photometric measurements in this
paper supplement this catalog, and provide a most comprehensive and homogeneous photometric
catalog for M31 GCs in the $JHK_{\rm s}$ bandpasses. Based on detailed comparisons,
our photometry is in good agreement with previous measurements.
Position (right ascension and declination) for cluster H26 is corrected here
(00:59:27.48 and +37:41:31.37).

We presented some statistical analysis based on the sample clusters, combined
with the photometric results of 10 GCs in the M31 halo \citep{ma12b}. We used
least square and maximum likelihood methods to fit a Gaussian and $t_5$ distribution on the
extinction-corrected $JHK_{\rm s}$ data.
The LF peaks for all the sample clusters derived by
fitting a $t_5$ distribution using maximum likelihood method are:
$J_0 = 15.726_{-0.190}^{+0.182}$, $H_0 = 15.095_{-0.182}^{+0.174}$, and ${K_{\rm s}}_0 = 14.816_{-0.144}^{+0.140}$
while for the confirmed GCs, the LF peaks are
$J_0 = 15.348_{-0.208}^{+0.206}$, $H_0 = 14.703_{-0.180}^{+0.176}$, and ${K_{\rm s}}_0 = 14.534_{-0.146}^{+0.142}$,
which are in good agreement with previous studies.
With the division of the confirmed GCs by both galactocentric
distance and metallicity, we found that the GCLFs are different between MR and MP,
inner and outer subpopulations. Generally, MP clusters are fainter than their MR counterparts,
and the inner clusters are brighter than the outer ones, which confirm previous results.

The NIR colors of the GC candidates are on average redder than those of the confirmed GCs,
which lead to an obscure bimodal distribution of the color indices.
The $(V-K_{\rm s})_0$ color index shows clearer correlation with [Fe/H] than other NIR colors,
indicating that those NIR intrinsic colors are less sensitive to metallicity.
The relation of $(V-K_{\rm s})_0$ and metallicity shows an obvious departure from linearity,
with a shallower slope towards the redder end.
The CMD and color-color diagram show that many GC candidates are located out of the evolutionary
tracks, indicating that some of them may not be true GCs.
The CMD also shows that the initial mass of M31 clusters covers a large range, and the
majority of the clusters have initial masses between $10^3$ and $10^6$ $M_{\odot}$.

\section*{Acknowledgments}

We especially thank the anonymous referee for his/her thorough report and helpful comments
and suggestions that have significantly improved the paper. This study has been supported by
the Chinese National Natural Science Foundation through grants 11373035, 11373033, and 11073032,
and by the National Basic Research Program of China (973 Program), No. 2014CB845702, 2014CB845704,
2013CB834902, and by the joint fund of Astronomy of the National Nature Science Foundation of
China and the Chinese Academy of Science, under Grants U1231113.

This publication makes use of data products from the Two Micron All Sky Survey, which is a
joint project of the University of Massachusetts and the Infrared Processing and Analysis
Center/California Institute of Technology, funded by the National Aeronautics and Space
Administration and the National Science Foundation.

We would like to thank the Digitized Sky Survey (DSS), which was produced at the Space
Telescope Science Institute under U.S. Government grant NAG W-2166. The images of DSS
are based on photographic data obtained using the Oschin Schmidt Telescope on Palomar
Mountain and the UK Schmidt Telescope.

\clearpage

\begin{table}
\begin{center}
\caption{Integrated Measurements for 913 Clusters in M31.}
\vspace{2mm} \label{t1.tab}
\begin{tabular}{lcccccccc}
\tableline
Name           & $J$               & $H$               & $K_{\rm s}$        & $r_{\rm ap}$   & $E(B-V)$  & Ref.$^a$  & [Fe/H]$_C$ & [Fe/H]$_K$  \\
               & (mag)             & (mag)             & (mag)             & ($''$)        & (mag)     &           & (dex)      & (dex)       \\
(1)            & (2)               & (3)               & (4)               & (5)          & (6)       & (7)       & (8)        & (9)         \\
\hline
AU008          & 15.218$\pm$0.269  & 14.403$\pm$0.265  & 14.299$\pm$0.324  & 4.0               & 0.21      & 5         & 99.99     & 99.99     \\
AU010          & 15.440$\pm$0.211  & 14.665$\pm$0.230  & 14.504$\pm$0.251  & 4.0               & 0.22      & 1         & -0.50     & 99.99     \\
B001           & 14.557$\pm$0.024  & 13.891$\pm$0.030  & 13.696$\pm$0.033  & 7.0               & 0.39      & 1         & -0.70     & -0.42     \\
B001D          & 17.369$\pm$0.107  & 17.435$\pm$0.238  & 17.334$\pm$0.254  & 4.0               & 0.13      & 6         & 99.99     & 99.99     \\
B002           & 15.797$\pm$0.050  & 15.483$\pm$0.070  & 15.247$\pm$0.068  & 5.0               & 0.11      & 1         & -2.20     & 99.99     \\
B003           & 15.689$\pm$0.045  & 14.996$\pm$0.056  & 15.087$\pm$0.074  & 6.0               & 0.12      & 1         & -1.60     & -0.99     \\
B004           & 14.966$\pm$0.030  & 14.371$\pm$0.034  & 14.136$\pm$0.034  & 5.0               & 0.18      & 1         & -0.70     & -1.00     \\
B005           & 13.414$\pm$0.014  & 12.720$\pm$0.017  & 12.598$\pm$0.018  & 7.0               & 0.16      & 1         & -0.70     & -0.82     \\
B006           & 13.273$\pm$0.013  & 12.635$\pm$0.015  & 12.423$\pm$0.016  & 12.0              & 0.17      & 1         & -0.50     & -0.59     \\
B006D          & 17.078$\pm$0.128  & 16.311$\pm$0.165  & 15.883$\pm$0.166  & 4.0               & 0.33      & 1         & 99.99     & -1.83     \\
\tableline
\end{tabular}
\end{center}
$^a$The reddening values are from: \citet{cald09, cald11} (ref=1),
\citet{fan08} (ref=2), \citet{bh00} (ref=3), \citet{fan10} (ref=4), mean reddening value of
clusters located within an annulus at every 2 kpc radius from the center of M31 (ref=5), and
foreground reddening value of $E(B-V)$ = 0.13 mag (ref=6).
\end{table}

\begin{table}
\begin{center}
\small
\caption{Clusters with Large Magnitude Scatters between This Paper and Previous Studies.}
\vspace{3mm} \label{t2.tab}
\begin{tabular}{lcccc}
\tableline
Name           & $\Delta J$   & $\Delta H$   & $\Delta K_{\rm s}$      & $r_{\rm ap}$ \\
               & (mag)        & (mag)        & (mag)                  & ($''$)          \\
(1)            & (2)          & (3)          & (4)                    & (5)          \\
\hline
\multicolumn{5}{c}{Comparison with 2MASS PSC}\\
\hline
B038D          & -1.10        &              &                        & 4            \\
B072D          & 1.611        &              &                        & 2            \\
B104           &              &              & -2.08                  & 4            \\
SK047B         & -1.16        &              &                        & 4            \\
SK049C         & -1.32        &              &                        & 4            \\
SK054C         &              &              & -3.36                  & 4            \\
SK086C         &              & 1.229        &                        & 4            \\
SK115B         &              &              & -1.40                  & 4            \\
SK136C         & 4.180        &              &                        & 4            \\
\hline
\multicolumn{5}{c}{Comparison with \citet{bh00,bhb01}}\\
\hline
DAO69          & 1.226        &              & 1.287                  & 4            \\
\hline
\multicolumn{5}{c}{Comparison with RBC V.5}\\
\hline
B017D          & 1.070        &              &                        & 5            \\
B041           &              & 1.551        & 1.974                  & 3            \\
B051D          & 1.411        &              &                        & 4            \\
B081D          & 1.366        &              &                        & 5            \\
B090           &              &              & -2.60                  & 4            \\
B104           &              & -1.06        &                        & 4            \\
B142D          &              &              & 1.077                  & 4            \\
B216           & 1.112        &              &                        & 4            \\
B244           & 1.153        &              &                        & 5            \\
B281D          & 1.153        &              &                        & 4            \\
B291D          & 1.104        &              &                        & 4            \\
B363           & 1.063        &              &                        & 5            \\
B452           & 1.264        &              &                        & 4            \\
B462           & 1.200        &              &                        & 4            \\
BA11           & 1.026        &              &                        & 4            \\
H7             &              &              & 1.164                  & 5            \\
M070           & 1.138        &              &                        & 5            \\
M091           &              & -1.52        &                        & 4            \\
NB16           & 1.050        &              &                        & 4            \\
SK049C         & -1.29        &              &                        & 4            \\
SK054C         &              &              & -3.28                  & 4            \\
SK213B         &              &              & 1.118                  & 6            \\
\tableline
\end{tabular}
\end{center}
\end{table}

\begin{table}
\renewcommand{\arraystretch}{1.2}
\centering
\caption{M31 GCLF Parameters.}
\vspace{3mm} \label{t3.tab}
\small
\tabcolsep=4.pt
\begin{tabular}{cccccccccc}
\tableline
\multicolumn{1}{c}{\multirow{1}{*}{Bandpass}}&\multicolumn{4}{c}{Least Square} &
\multicolumn{4}{c}{Maximum Likelihood} & \multicolumn{1}{c}{\multirow{1}{*}{$N$}} \\
\cmidrule(r){2-5}\cmidrule(l){6-9}
          & $\mu_{G}$ & $\sigma_{G}$ & $\mu_{t}$ & $\sigma_{t}$& $\mu_{G}$ & $\sigma_{G}$ & $\mu_{t}$    & $\sigma_{t}$ &  \\
          & (mag)   & (mag)        & (mag)   & (mag)        & (mag)   & (mag)        & (mag)         & (mag)        &  \\
(1)       & (2)     & (3)          & (4)     & (5)          & (6)     & (7)          & (8)           & (9)          &  (10)\\
\hline
\multicolumn{9}{c}{All sample clusters}\\
\hline
$J_0$     &$ 15.873_{-0.198}^{+0.190}$       &$ 1.406_{-0.142}^{+0.176}$        &$ 15.893_{-0.192}^{+0.184}$       &$ 1.355_{-0.138}^{+0.174}$        &$ 15.726_{-0.198}^{+0.190}$       &$ 1.393_{-0.142}^{+0.178}$        &$ 15.726_{-0.190}^{+0.182}$       &$ 1.393_{-0.132}^{+0.170}$     &   914       \\
$H_0$     &$ 15.224_{-0.186}^{+0.178}$       &$ 1.340_{-0.136}^{+0.170}$        &$ 15.245_{-0.184}^{+0.174}$       &$ 1.290_{-0.132}^{+0.168}$        &$ 15.095_{-0.186}^{+0.178}$       &$ 1.377_{-0.136}^{+0.170}$        &$ 15.095_{-0.182}^{+0.174}$       &$ 1.377_{-0.130}^{+0.166}$     &   892       \\
${K_{\rm s}}_0$&$ 14.887_{-0.144}^{+0.140}$       &$ 1.237_{-0.108}^{+0.134}$        &$ 14.903_{-0.146}^{+0.140}$       &$ 1.198_{-0.106}^{+0.134}$        &$ 14.816_{-0.144}^{+0.140}$       &$ 1.324_{-0.108}^{+0.136}$        &$ 14.816_{-0.144}^{+0.140}$       &$ 1.324_{-0.104}^{+0.132}$    &    872       \\
\hline
\multicolumn{9}{c}{Confirmed GCs}\\
\hline
$J_0$     &$ 15.405_{-0.222}^{+0.218}$       &$ 1.505_{-0.160}^{+0.196}$        &$ 15.409_{-0.212}^{+0.208}$       &$ 1.459_{-0.148}^{+0.192}$        &$ 15.348_{-0.220}^{+0.218}$       &$ 1.424_{-0.158}^{+0.196}$        &$ 15.348_{-0.208}^{+0.206}$       &$ 1.423_{-0.144}^{+0.184}$    &    579       \\
$H_0$     &$ 14.816_{-0.192}^{+0.186}$       &$ 1.431_{-0.134}^{+0.170}$        &$ 14.829_{-0.182}^{+0.178}$       &$ 1.387_{-0.128}^{+0.164}$        &$ 14.703_{-0.192}^{+0.186}$       &$ 1.367_{-0.134}^{+0.168}$        &$ 14.703_{-0.180}^{+0.176}$       &$ 1.367_{-0.124}^{+0.158}$    &    565       \\
${K_{\rm s}}_0$&$ 14.621_{-0.150}^{+0.144}$       &$ 1.354_{-0.108}^{+0.136}$        &$ 14.640_{-0.148}^{+0.142}$       &$ 1.310_{-0.106}^{+0.134}$        &$ 14.534_{-0.150}^{+0.144}$       &$ 1.357_{-0.108}^{+0.136}$        &$ 14.534_{-0.146}^{+0.142}$       &$ 1.357_{-0.102}^{+0.132}$    &    559       \\
\hline
\multicolumn{9}{c}{Confirmed GCs (metallicity available from Kang et al.)}\\
\hline
$J_0$     &$ 14.583_{-0.148}^{+0.144}$       &$ 1.192_{-0.110}^{+0.134}$        &$ 14.596_{-0.144}^{+0.140}$       &$ 1.141_{-0.108}^{+0.132}$        &$ 14.534_{-0.146}^{+0.146}$       &$ 1.180_{-0.110}^{+0.134}$        &$ 14.534_{-0.144}^{+0.138}$       &$ 1.180_{-0.104}^{+0.130}$    &    287       \\
$H_0$     &$ 14.031_{-0.132}^{+0.130}$       &$ 1.214_{-0.098}^{+0.120}$        &$ 14.051_{-0.130}^{+0.126}$       &$ 1.162_{-0.096}^{+0.122}$        &$ 13.982_{-0.134}^{+0.130}$       &$ 1.221_{-0.098}^{+0.122}$        &$ 13.982_{-0.130}^{+0.126}$       &$ 1.214_{-0.094}^{+0.120}$     &   287       \\
${K_{\rm s}}_0$&$ 13.902_{-0.100}^{+0.098}$       &$ 1.214_{-0.074}^{+0.088}$        &$ 13.927_{-0.098}^{+0.096}$       &$ 1.150_{-0.076}^{+0.090}$        &$ 13.862_{-0.100}^{+0.098}$       &$ 1.225_{-0.074}^{+0.088}$        &$ 13.861_{-0.098}^{+0.094}$       &$ 1.161_{-0.074}^{+0.088}$   &     288       \\
\tableline
\end{tabular}
\end{table}

\begin{table}
\renewcommand{\arraystretch}{1.2}
\centering
\caption{M31 GCLF Parameters for Four Groups of the Confirmed GCs.}
\vspace{3mm} \label{t4.tab}
\tabcolsep=4.pt
\begin{tabular}{cccccccccc}
\tableline
\multicolumn{1}{c}{\multirow{1}{*}{Subsample}}&\multicolumn{3}{c}{$J_0$} &
\multicolumn{3}{c}{$H_0$} & \multicolumn{3}{c}{${K_{\rm s}}_0$} \\
\cmidrule(l){2-4}\cmidrule(l){5-7}\cmidrule(l){8-10}
          & $\mu_{t}$& $\sigma_{t}$ & $N$     & $\mu_{t}$& $\sigma_{t}$ &  $N$     & $\mu_{t}$   & $\sigma_{t}$ & $N$   \\
          & (mag)   & (mag)        &         & (mag)   & (mag)        &          & (mag)        & (mag)     &    \\
(1)       & (2)     & (3)          & (4)     & (5)     & (6)          & (7)      & (8)          & (9)       & (10)\\
\hline
MR inner  &$ 14.140_{-0.120}^{+0.118}$       &$ 1.157_{-0.084}^{+0.104}$        & 139     &$ 13.548_{-0.098}^{+0.096}$       &$ 1.171_{-0.070}^{+0.084}$        & 139     &$ 13.400_{-0.110}^{+0.106}$       &$ 1.130_{-0.076}^{+0.092}$        & 139     \\
MP inner  &$ 14.904_{-0.064}^{+0.064}$       &$ 1.067_{-0.054}^{+0.064}$        & 83      &$ 14.413_{-0.064}^{+0.062}$       &$ 1.123_{-0.056}^{+0.066}$        & 84      &$ 14.294_{-0.050}^{+0.050}$       &$ 1.022_{-0.044}^{+0.048}$        & 84      \\
MR outer  &$ 14.672_{-0.044}^{+0.044}$       &$ 0.874_{-0.032}^{+0.036}$        & 25      &$ 14.118_{-0.062}^{+0.064}$       &$ 0.980_{-0.042}^{+0.050}$        & 25      &$ 13.910_{-0.066}^{+0.064}$       &$ 0.996_{-0.046}^{+0.054}$        & 25      \\
MR outer  &$ 15.033_{-0.080}^{+0.084}$       &$ 1.079_{-0.058}^{+0.068}$        & 43      &$ 14.484_{-0.052}^{+0.052}$       &$ 1.041_{-0.036}^{+0.042}$        & 42      &$ 14.430_{-0.044}^{+0.046}$       &$ 1.009_{-0.034}^{+0.038}$        & 43      \\
\tableline
\end{tabular}
\end{table}

\begin{table}
\renewcommand{\arraystretch}{1.2}
\begin{center}
\caption{Results from the KMM Homoscedastic Bimodality Tests for the $JHK_{\rm s}$ Colors of
the Sample Clusters in M31.}
\vspace{3mm} \label{t5.tab}
\begin{tabular}{lcccccccc}
\tableline
Data set  & $\overline{\Delta}$  & $\sigma_{\Delta}$ & $\overline{\Delta_1}$ & $\overline{\Delta_2}$ & $\sigma_{\Delta}$ &  $N_1$  &  $N_2$  &   $p$  \\
                & (mag)          & (mag)           & (mag)                & (mag)                  & (mag)           &         &         &        \\
(1)             & (2)            & (3)             & (4)                  & (5)                    & (6)             & (7)     & (8)     & (9)    \\
\hline
$(J-H)_0$          & 0.600     & 0.101   & 0.579   & 1.573   & 0.080   & 840     & 16      & 0.000     \\
$(H-K_{\rm s})_0$   & 0.240     & 0.114   & 0.193   & 0.979   & 0.079   & 779     & 38      & 0.000     \\
$(J-K_{\rm s})_0$   & 0.845     & 0.155   & 0.801   & 1.622   & 0.120   & 812     & 29      & 0.000     \\
$(V-K_{\rm s})_0$   & 2.423     & 0.696   & 2.324   & 2.560   & 0.683   & 718     & 125     & ...       \\
\tableline
\end{tabular}
\end{center}
\end{table}

\begin{table}
\renewcommand{\arraystretch}{1.2}
\begin{center}
\caption{Results from the KMM Homoscedastic Bimodality Tests for the Metallicities of the Sample
Clusters in M31.}
\vspace{3mm} \label{t6.tab}
\begin{tabular}{lcccccccc}
\tableline
Data set  & $\overline{\rm [Fe/H]}$ & $\sigma_{\rm [Fe/H]}$ & $\overline{\rm [Fe/H]_1}$ & $\overline{\rm [Fe/H]_2}$ & $\sigma_{\rm [Fe/H]}$ &  $N_1$  &  $N_2$   &   $p$ \\
               & (dex)              & (dex)               & (dex)                     & (dex)                    & (dex)               &         &         &        \\
(1)            & (2)                & (3)                 & (4)                       & (5)                      & (6)                 & (7)     & (8)     & (9)    \\
\hline
Caldwell et al. &   -1.010          &  0.345              &  -0.934                   &  -1.772                  &  0.287              &  303    &  9      &  0.361 \\
Kang et al.     &   -1.032          &  0.350              &  -0.632                   &  -1.527                  &  0.152              &  167    &  127    &  0.002 \\
\tableline
\end{tabular}
\end{center}
\end{table}

\clearpage
\begin{center}
{APPENDIX}
\end{center}
\begin{center}
{SEVERAL SANITY CHECKS OF THE 2MASS PHOTOMETRY}
\end{center}

This appendix addresses the need to clarify the procedure adopted to 
obtain the best measurement of the aperture photometry of each GC candidate.
Two checks of the 2MASS photometry were given as follows.

A test of the photometry comparing results from the 2MASS-6X 
and the 2MASS All-Sky images was performed,
considering that objects with low S/N may have underestimated magnitudes being near 
the magnitude limits of the 2MASS All-Sky images. 
This test may shed some light on the amount of photons lost for faint objects and 
provide the safest magnitude limits at which to stop the analysis.
Figure \ref{fig1} displays the photometry comparison of the same objects in both 2MASS-6X and 
2MASS All-sky images, with same photometry methods. 
However, no clear trend can be seen that photons are lost for faint objects.
In this paper, no magnitude limits were set to stop the statistical analysis.

\begin{figure}[htp!]
\center
{\includegraphics[width=0.8\textwidth]{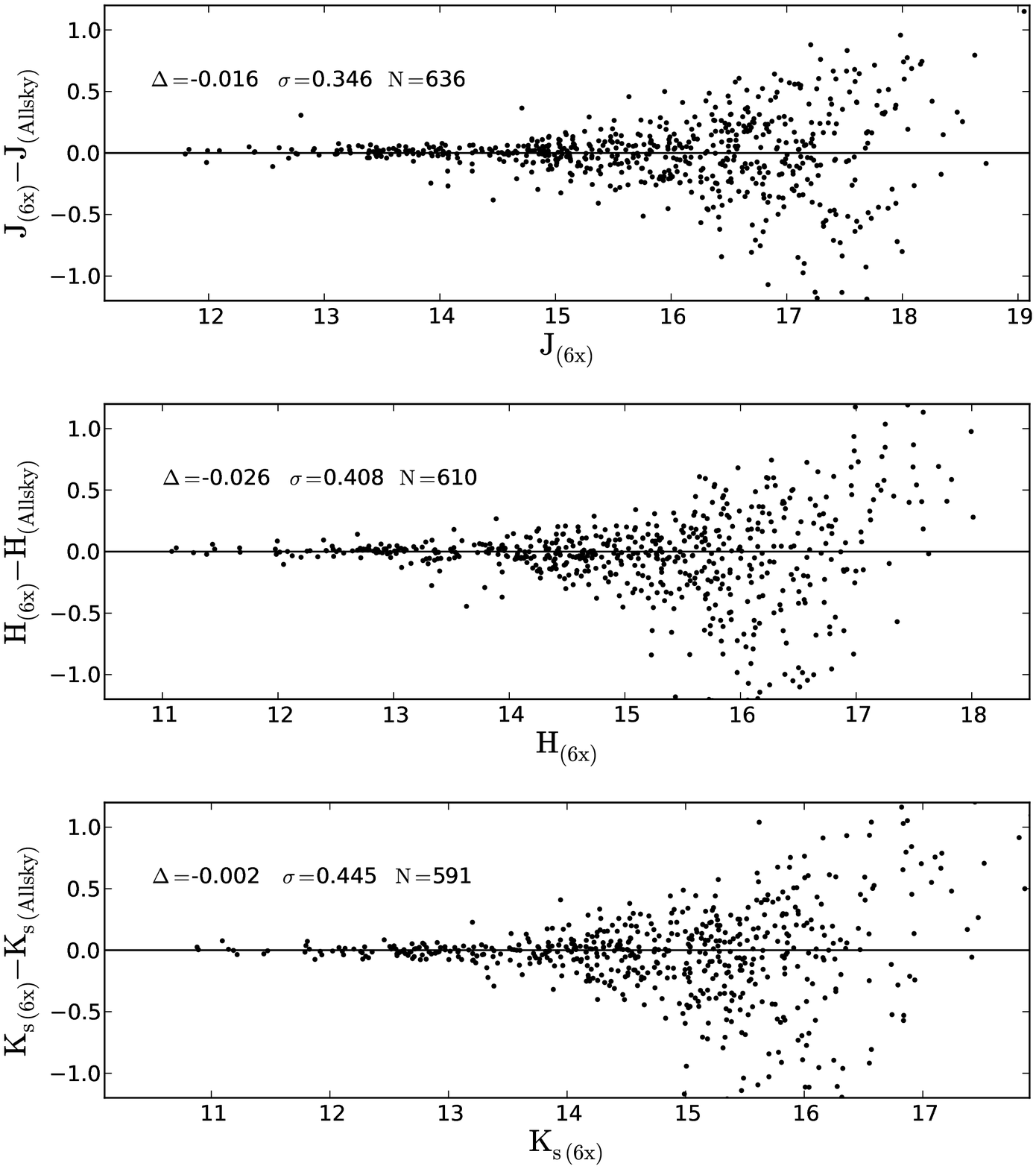}}
\vspace{-0.7cm}
\caption{Comparison of the obtained photometry from the 2MASS-6X and the 2MASS All-sky images.}
\label{fig1}
\end{figure}

\begin{figure}[htp!]
\center
{\includegraphics[width=0.8\textwidth]{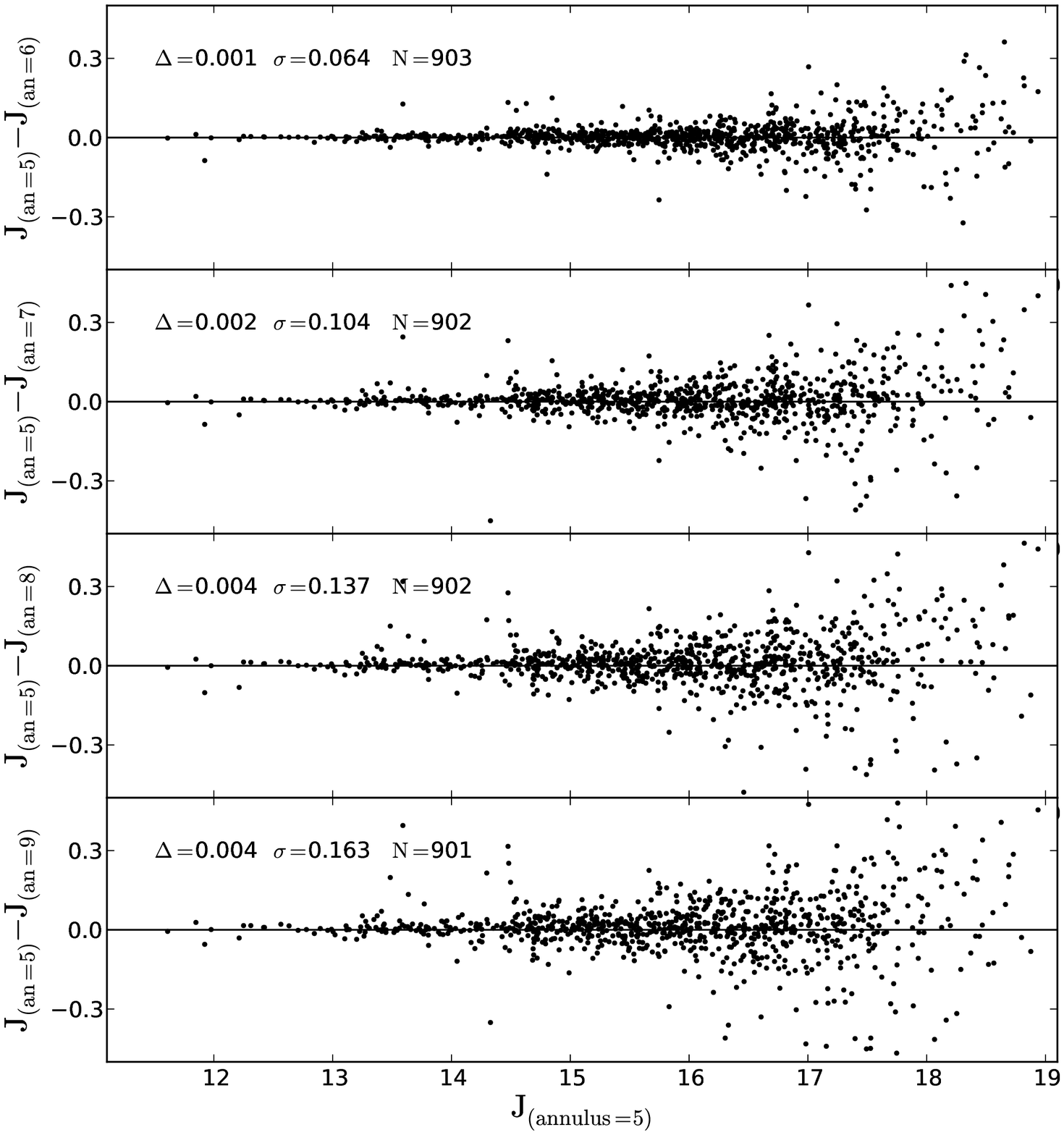}}
\vspace{-0.7cm}
\caption{Comparison of the obtained $J$-band photometry with different set of annulus.}
\label{fig2}
\end{figure}

As was mentioned in section \ref{phot.sec}, 
we choosed the annulus, which is the ``dannulus'' in {\sc daophot}, 
of 5 pixels following our experience.
A test about the effect of annulus on photometry was perfromed, with annulus
ranging from 5 pixels to 9 pixels. 
Figure \ref{fig2} shows the comparison of $J$-band magnitudes with different set of annulus.
No significant trend can be seen.
We think that the annulus set in this paper is reasonable and 
could assure a proper estimate of the sky. 

\end{document}